# On the Use of Letters of Recommendation in Astronomy and Astrophysics Graduate Admissions

**Darcy Barron[1] Rachel Bezanson[2] Laura Blecha[3] Laura Chomiuk[4] Lia Corrales[5] Vera Gluscevic[6] Kristen McQuinn[7] Anne Medling[8] Noel Richardson[9] Ryan Trainor[10] Jessica Werk[11]**

[1]University of New Mexico, [2]University of Pittsburgh, [3]University of Florida, [4]Michigan State University, [5]University of Michigan, [6]University of Southern California, [7]Rutgers University, [8]University of Toledo, [9]Embry-Riddle Aeronautical University, [10]Franklin & Marshall College, [11]University of Washington

## Abstract

Letters of recommendation are a common tool used in graduate admissions. Most admissions systems require three letters for each applicant, burdening both letter writers and admissions committees with a heavy work load that may not be time well-spent. Most applicants do not have three research advisors who can comment meaningfully on research readiness, adding a large number of letters that are not useful. Ideally, letters of recommendation will showcase the students' promise for a research career, but in practice, the letters often do not fulfill this purpose.

As a group of early and mid-career faculty who write dozens of letters every year for promising undergraduates, we are concerned and overburdened by the inefficiencies of the current system. In this open letter to the AAS Graduate Admissions Task Force, we offer an alternative to the current use of letters of recommendation: a portfolio submitted by the student, which highlights e.g., a paper, plot, or presentation that represents their past work and readiness for grad school, uploaded to a centralized system used by astronomy and astrophysics PhD programs. While we argue that we could eliminate letters in this new paradigm, it may instead be advisable to limit the number of letters of recommendation to one per applicant.

## What is the intended purpose of recommendation letters? Graduate admissions in the ideal world

Applying to a PhD program is a crucial stage in a young astronomer's career development. Typical graduate school applications in our field require student-developed statements on personal goals, research history, and life experiences. In addition to this, students are usually required to submit a curriculum vitae, official university transcripts, and three letters of recommendation from advisors and/or instructors. Through this application packet, admissions committees hope to learn enough about the applicant to appropriately assess how well each student would fit into and succeed in their program.

There are various qualities that an admissions committee might value that are not easily assessed by transcripts and resumes. Such qualities include "soft skills" (like professionalism, teamwork, time management, and communication) and non-cognitive factors (like grit, motivation, and resilience). There can also be a need to place the student in context, in order to better understand the quirks of how a student's background has prepared them for grad school, the student's stand-out qualities, and how they rank relative to their peers. An appropriate reference letter can provide valuable insight into many of these qualities from the perspective of a faculty peer. They can also serve a critical role in promoting equity in graduate admissions. For example, the level of ownership and preparedness that a student might claim strongly reflects their level of confidence. Minoritized students (for example: women, students of color, LGBTQ+ students, and students with disabilities) may feel less confident and might therefore adopt a passive tone that understates their contributions and competencies (Osman et al. 2015). Letters of reference can provide a valuable external assessment of a student's skills, independent of their confidence level. Effective, well-written, and strategic letters of recommendation can also bridge the experience gap of students from under-resourced institutions by highlighting personal characteristics that could make them successful in PhD research. Useful letters can also help to illuminate any special circumstances that affected the student's academic track record, alleviate or raise any potential concerns about future performance, and describe hidden talents (e.g., leadership and creativity) that a student may not recognize in themselves.

While a letter of recommendation has the potential to serve all of the above objectives, we explain below how their role in the current framework of graduate admissions has fallen short of their intended purpose. As we will argue in this letter, changing the design of a graduate application packet to no longer rely on letters of recommendation, and to instead consider a student portfolio and context statements from the advisors specific to the portfolio products, could bring us closer to a more equitable and effective admissions process.

## Who are we? and Why are we writing this?

Our collaboration was motivated to write this open letter to the Graduate Admissions Task Force, convened by the American Astronomical Society (AAS), over discussions at the 2024 Cottrell Scholars Conference, which is an annual conference funded by the Research Corporation for Science Advancement (RCSA). Cottrell Scholars are tenure-stream faculty with a demonstrated interest in undergraduate education and strong scientific records. We are funded by individual and collective RCSA grants, including a current Cottrell Scholars Collaborative grant. We include amongst our ranks faculty members at primarily undergraduate-serving institutions and faculty at more research-intensive institutions. The authors of this paper actively advise undergrads in research (some of us with large groups of 10–20 students), and are deeply invested in their success. Some of us are undergraduate advisors or undergraduate program coordinators for astrophysics majors at our universities. Many of us sit on graduate admissions committees at large R1/R2 universities where the challenges of grad admissions are well understood.

The AAS Graduate Admissions Task Force has discussions underway throughout 2024, with the goal of providing "findings and recommendations to the AAS Board of Trustees on if/how our community can mitigate the stress of graduate admissions on both departments and applicants."

## What purposes do letters currently serve, practically?

A record number of students are applying to astronomy and astrophysics Ph.D. programs, with programs regularly receiving hundreds of applications every admissions cycle for fewer than 20 admissions offers. Because of these increasingly unfavorable odds, many students apply to 10–20 programs, each requiring a standard three letters of recommendation. Admissions committees, often consisting of only handful of faculty members, are therefore tasked with reading hundreds, if not more than 1000 letters, from their colleagues.

The research advisors and instructors, asked to write 1–2 page letters evaluating the student's overall potential for success in graduate school, are further burdened by navigating a multitude of admissions portals with disparate ranking scales and technical designs. Faculty who are advising the most undergraduates in research or teaching seminal advanced undergraduate courses can have more than 10 students every year for whom they are advocating. This process requires an enormous investment of faculty time — from both the committee perspective and the letter-writer perspective. It is important to ask ourselves if this investment is well spent.

Under these conditions, admissions committee members must cursorily skim letters of recommendation on their first-pass application reads. Honing in on carefully worded recommendations with thoughtful assessments of students' future potential for success in a Ph.D. program is difficult, as high-quality letters are lost in a sea of more superficial recommendations, driven by the requirement that each student have three letters (we have noticed that even the most promising, active students rarely have deep relationships with three potential letter writers, and often need to request a letter from e.g., an instructor who has only taught them in a single class). The proliferation of Likert-like scale rankings assessing a variety of student qualities with a range of scales—different for every graduate program—further muddies the water and propagates bias. For an admissions committee hoping to evaluate students fairly and equitably, there is no practical way to spend enough time reading these letters for them to serve their intended purpose. This heavy workload is in itself bad for equity in graduate admissions, as research shows that when evaluation is hasty, evaluators (be they the letter writer or the admissions committee) are more likely to fall back on schema and stereotypes (Devine 1989, Morgan et al. 2013).

The mere act of requiring each student to identify three letter writers already biases graduate admission against students from smaller or less-resourced institutions. Many of the faculty writing these letters do not serve on graduate committees and receive little guidance about how these letters and ranking scales are evaluated in practice. For example, letters from course instructors discussing students' classwork and in-class participation are widely ignored and carry little weight. Conversely, students who have had the opportunity to participate in multiple research projects will be at an advantage (given the current 3-letter requirement). Meanwhile, letters from prominent scientists or collaborators known to the admissions committee are often given outsized weight, whether intentionally or unintentionally. This latter use exacerbates existing inequities based on an applicant's access to resources at their institutions. **Therefore, while the context provided by a faculty research advisor who knows the student well can be valuable to the admissions process, it is worthwhile to consider whether all of the packaging of the typical three letters of recommendation is effective and efficient at accomplishing this goal.**

Based on conversations with members of graduate admissions committees at our and other institutions, we understand that letters are an aspect of the application that are regarded in a variety of ways. Instead of serving idealized purposes, these letters most commonly serve the practical purposes of:

- Searching for red flags indicating a student's lack of preparation for graduate school
- Finding a strong letter from a collaborator or scientist known to and trusted by an admissions committee member

- Contextualizing the students' contribution to the original research they list on their CVs and discuss in their statements

However, letters of recommendation are often not effective at meeting these goals, and do not sufficiently satisfy admissions committees' need to understand what the student accomplished during prior research experiences. While letter writers attempt to speak to these matters, the combination of "letter inflation" (where every student walks on water) and the increasing use of vague platitudes to describe student work limit the usefulness of letters. Because students often need to "dig deep" to find three letter writers, many letter writers may not know the student well enough to accurately comment on the student's readiness for grad school. We are already hearing many reports of the usage of artificial intelligence and large language models to write reference letters, and this is likely only to increase in the future, as technology improves and if pressure on letter writers continues to build. Even when written by a human who knows the student, letters can miss the mark. It is a long-standing finding in the psychology literature that letters of recommendation have poor inter-rater reliability; for example, as summarized by Aamodt et al. (1993): "The reliability problem is so severe that Baxter et al. (1981) found that there is more agreement between two recommendations written by the same person for two different applicants than there is between two people writing recommendations for the same person".

The practical uses of letters of recommendation in admission processes further introduce a number of biases that are known to disadvantage students from under-represented backgrounds. Studies consistently find that the content of reference letters depends on the applicant's gender and race. For example, letters of reference describe white applicants as having more agency and competency than Black and Latinx applicants (Grimm et al. 2020, Rojek et al. 2019). Similar discrepancies are found with gender (Madera et al. 2009), with female applicants described with fewer "ability" words and more "grindstone" words (Schmader et al. 2007). Letter writers have been found to cast more doubt on the records of female and/or non-white applicants, compared to their male and/or white colleagues, despite similar grade point average distributions (Houser & Lemmons 2017, Madera et al. 2019). Fundamentally, bias will be unavoidable in any competitive application process, but every extraneous opinion or piece of "packaging" unnecessarily increases this potentially problematic noise.

## How is the current system of letters of recommendation failing?

In the current system, three letters is simply too many. As mentioned previously, most students will have one research advisor that can speak to their potential as a researcher in graduate school and beyond, meaning only one letter would carry any meaningful weight. Early career faculty, faculty at primarily undergraduate institutions, and faculty from underrepresented backgrounds are more likely to advise a large number of research students (Lin & Kennette 2022). The letter of recommendation process for these valuable letter writers often translates to more than a full work week over the admission season spent writing letters that will likely be skimmed, submitting those letters, and filling out online surveys.

In addition to the burden reference letters place on letter writers and admissions committees alike, reference letters are not meeting our goals for graduate admissions. Reference letter propagate current inequities in the system. Moreover, reference letters often lack the information admission committees seek, or admissions committees struggle to trust letters. Describing and contextualizing students' research accomplishments is a central objective of reference letters, but a reference letter on its own is often not sufficient evidence to demonstrate research prowess. Admissions committees crave direct evidence in the form of research products, but rarely are such products made available in the application, with the exception of published papers in the peer-reviewed literature. This leads to an "all or nothing" situation, where a student gets a huge boost in their application to grad school if they have managed to publish a paper, and they have little to show for their hard work if the paper does not get completed in time. In the following Section, we propose an alternative that will provide every student an opportunity to directly demonstrate their readiness for grad school: a portfolio.

## A proposed plan: A portfolio accompanied by a statement of context from a mentor/advisor

### The Portfolio

Here we present an alternative to reference letters, which enables the student to directly demonstrate skills relevant for graduate school while minimizing the involvement of faculty mentors: the portfolio.

A student's portfolio would consist of work products created by them, relevant to the research process; these could have been created during a research experience, or during e.g.,  a class or capstone project. Examples of products include:

- Poster — could be a conference poster (e.g., AAS), or something presented at a research symposium, or a departmental/university event
- Writing sample — could be from a class writing project, REU or capstone report, school newspaper, or blog post
- Code sample — could be from a class, github repo, or personal project
- Graphic — could be an exceptional plot or infographic, artwork, etc
- Journal article (in prep, submitted, or accepted); this can be a co-authored paper or a first-author paper.
- Recorded talk — a link to e.g., a youtube video of an oral presentation
- Design files or photos of something the student designed and/or made (e.g., CAD, STY, circuits)

As part of the portfolio, the student would highlight 1–2 of these products for grad admissions committees to focus on.

For each highlight, the student would write a paragraph explaining the context in which they created the product, their role in the work, and how it ties into their interest in pursuing a graduate education. It is acceptable that one or both of these highlights were prepared with the help of others, in a collaborative context, but the student must describe how they specifically contributed.

Portfolios should be hosted in a single, central location which might be facilitated by e.g., AAS or AIP. The student could decide if they would like to make the portfolio publicly viewable or if they would like to restrict access to the grad schools they have applied to.

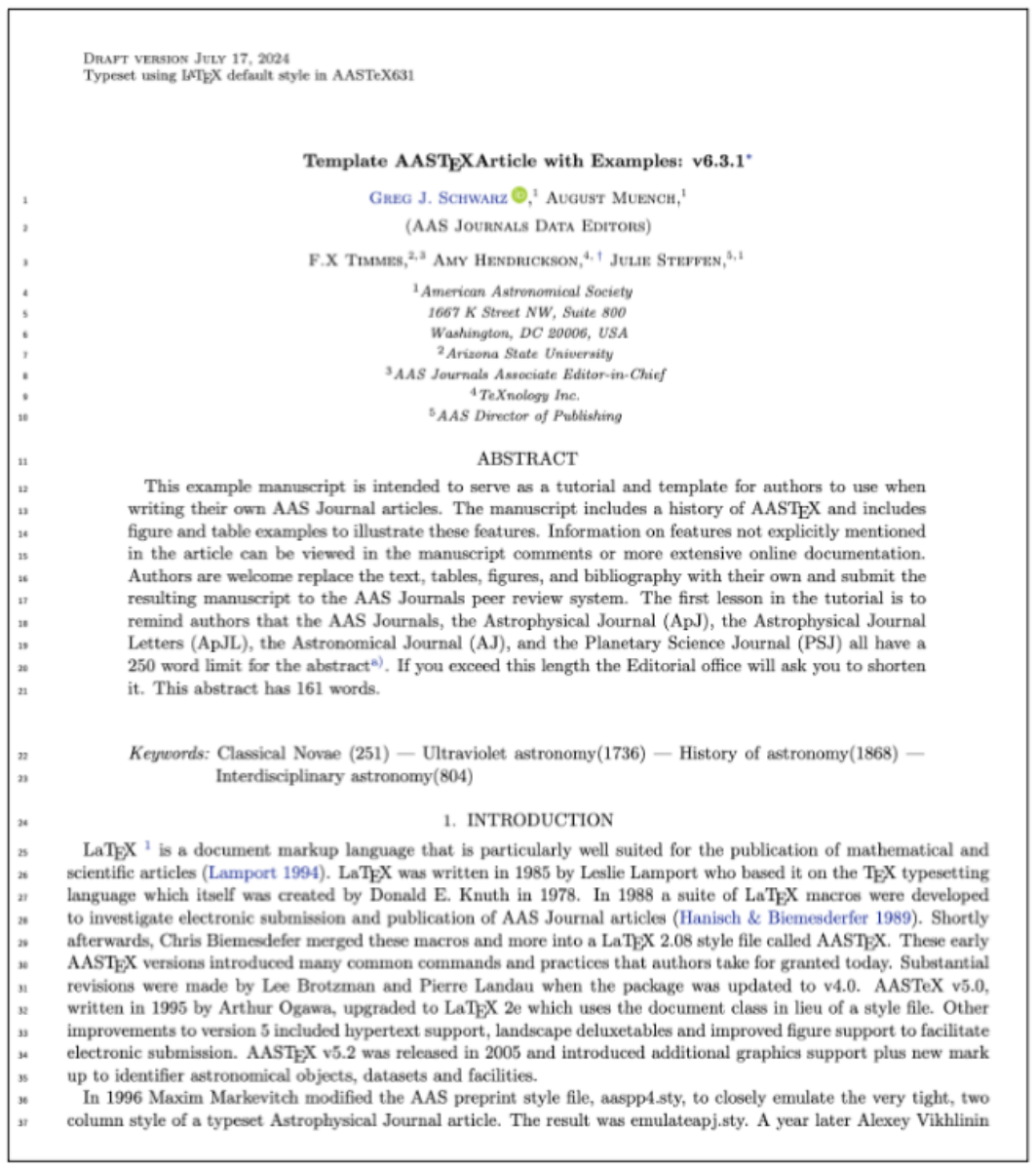
Draft version July 17, 2024
Typeset using LaTeX default style in AASTeX631

**Template AASTeXArticle with Examples: v6.3.1***

Greg J. Schwarz,[1] August Muench,[1]
(AAS Journals Data Editors)
F.X Timmes,[2,3] Amy Hendrickson,[4,†] Julie Steffen,[5,1]

[1] *American Astronomical Society*
*1667 K Street NW, Suite 800*
*Washington, DC 20006, USA*
[2] *Arizona State University*
[3] *AAS Journals Associate Editor-in-Chief*
[4] *TeXnology Inc.*
[5] *AAS Director of Publishing*

ABSTRACT

This example manuscript is intended to serve as a tutorial and template for authors to use when writing their own AAS Journal articles. The manuscript includes a history of AASTeX and includes figure and table examples to illustrate these features. Information on features not explicitly mentioned in the article can be viewed in the manuscript comments or more extensive online documentation. Authors are welcome replace the text, tables, figures, and bibliography with their own and submit the resulting manuscript to the AAS Journals peer review system. The first lesson in the tutorial is to remind authors that the AAS Journals, the Astrophysical Journal (ApJ), the Astrophysical Journal Letters (ApJL), the Astronomical Journal (AJ), and the Planetary Science Journal (PSJ) all have a 250 word limit for the abstract[a)]. If you exceed this length the Editorial office will ask you to shorten it. This abstract has 161 words.

*Keywords:* Classical Novae (251) — Ultraviolet astronomy(1736) — History of astronomy(1868) — Interdisciplinary astronomy(804)

1. INTRODUCTION

LaTeX [1] is a document markup language that is particularly well suited for the publication of mathematical and scientific articles (Lamport 1994). LaTeX was written in 1985 by Leslie Lamport who based it on the TeX typesetting language which itself was created by Donald E. Knuth in 1978. In 1988 a suite of LaTeX macros were developed to investigate electronic submission and publication of AAS Journal articles (Hanisch & Biemesderfer 1989). Shortly afterwards, Chris Biemesdefer merged these macros and more into a LaTeX 2.08 style file called AASTeX. These early AASTeX versions introduced many common commands and practices that authors take for granted today. Substantial revisions were made by Lee Brotzman and Pierre Landau when the package was updated to v4.0. AASTeX v5.0, written in 1995 by Arthur Ogawa, upgraded to LaTeX 2e which uses the document class in lieu of a style file. Other improvements to version 5 included hypertext support, landscape deluxetables and improved figure support to facilitate electronic submission. AASTeX v5.2 was released in 2005 and introduced additional graphics support plus new mark up to identifier astronomical objects, datasets and facilities.

In 1996 Maxim Markevitch modified the AAS preprint style file, aaspp4.sty, to closely emulate the very tight, two column style of a typeset Astrophysical Journal article. The result was emulateapj.sty. A year later Alexey Vikhlinin

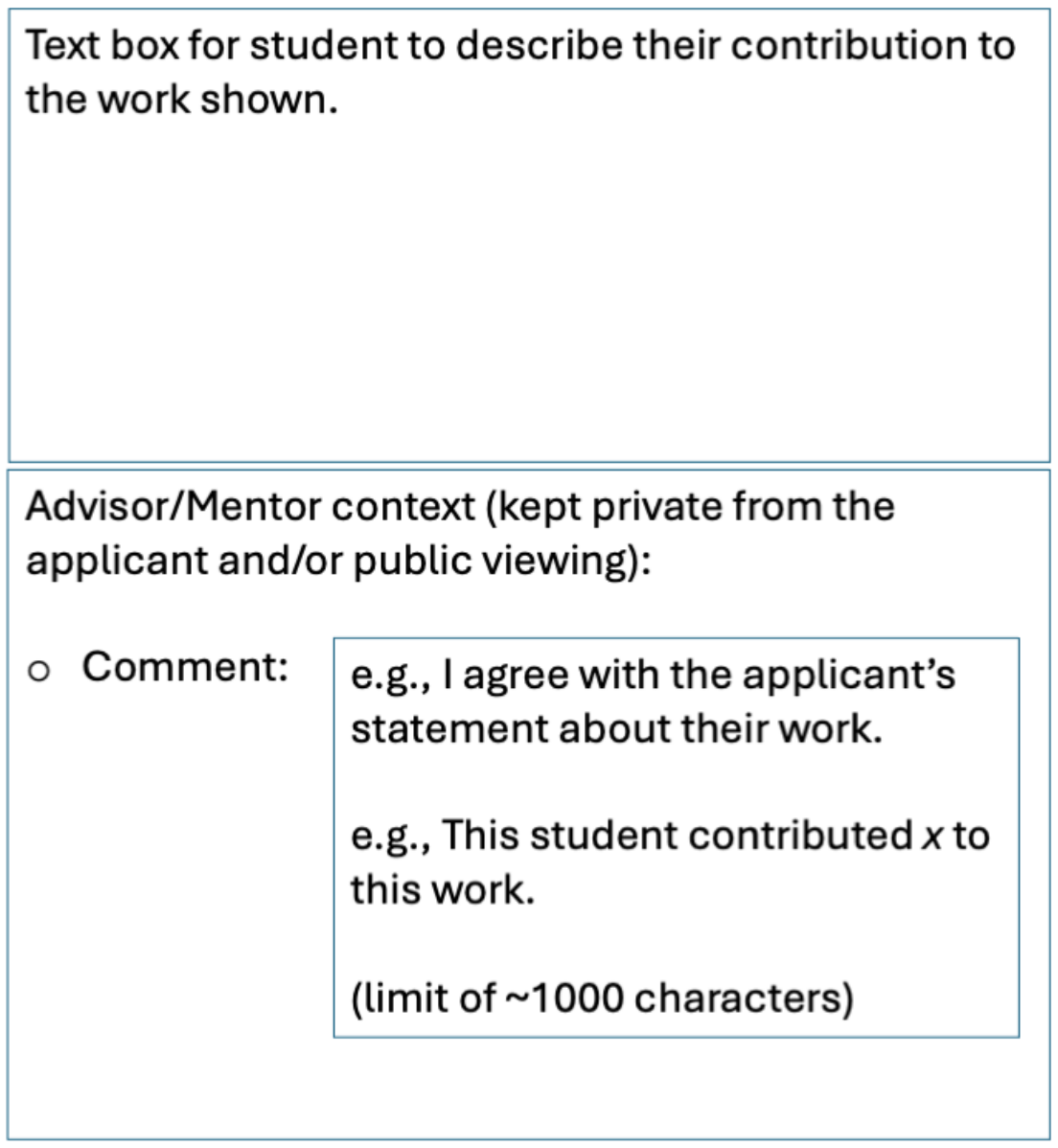

A mock-up of how a portfolio highlight could be uploaded to a central system.

### The Context Statement (submitted by an advisor/mentor)

To accompany the student's portfolio, for each highlight, the student will designate one advisor or mentor who will submit a brief statement to further contextualize and corroborate the student's work. The context statement should assess whether the student's description of their contributions is accurate, and may provide additional information on the student's strengths and/or weaknesses that were demonstrated in preparing the highlight.

Each context statement should be brief, for example, limited to 1000 characters. If the student highlights two products, different advisors may be designated for each. The context statement is uploaded to a single, central location where the portfolio is hosted, and is viewable by only those schools to which the student applies (and not the student).

### How the Portfolio/Context Statements Improve Grad Admissions

Our collaboration believes the combination of the student portfolio and advisor context statements are preferable to the status quo of three recommendation letters for the following reasons:

- The portfolio and advisor context statements will be better for equity. The portfolio is an appreciation of the diversity of student work, and that not everyone has access to a long-term research project or a research mentor who will shepherd them through writing a journal article. It avoids the "all or nothing" current state, where students either are an author on a refereed paper or have no research products to share. Students can share results from relatively modest/short-term projects, or results carried out in class. Prowess in e.g., critical thinking, creativity, scientific communication, data visualization, etc. can be demonstrated without the need for in-depth, long-term immersion in a research project.
- Our group believes that the specificity of the prompt for a context statement from the advisor will cut down on the likelihood of bias or subjectivity of letters or recommendation, making all applications similar in their focus. It also makes the context statement more concise, cutting down on the noise encountered by admissions committees.
- The portfolio and advisor context statements will be uploaded to a central database to allow for uniformity across institutions that the students apply to, and creating less work for their faculty references.
- The burden on faculty letter writers will be greatly lessened, with concise context statements requested from 1–2 references, rather than full-length letters from 3 references. The overhead of uploading/reviewing the student on many diverse admissions portals will be removed, as the portfolio and context statements will be housed in a single, centralized location.

## Potential Challenges

We foresee a few challenges that could become apparent with the implementation of the portfolio instead of letters, which include:

1. Hosting the central repository, with context statements that may include personal or health information about the student, may have challenges with student privacy (e.g., FERPA and/or HIPAA).
2. Hosting the central repository might be expensive and could require cost-sharing between universities.
3. The content of context statements may end up with similar biases and problems associated with letters of reference, even if kept to a strict short character limit and a focused purpose.
4. Product-based context statements do not provide an opportunity for references to explain other aspects of the student's application package: for example, personal challenges leading to a gap in schooling or low GPA, and how the student overcame them.
5. Students may want to highlight different items depending on the specific program that they are applying to.
6. Reference letters may be required for university fellowships; every university is likely to have different rules on this.
7. It may prove difficult and time-consuming for admissions committees to assess the portfolio highlights. Still, this effort will be spent reviewing products that directly tie to the students' records and abilities, rather than spending valuable committee time parsing subjective letters.

In the process of a new system being adopted, there is likely to be a long period of overlap where faculty and graduate admissions committees will have the burden of both systems. During this time, a new system needs to have a compelling advantage with broad appeal to overcome the inertia of the existing system.

## Conclusions

In this open letter, we have described some of the challenges and inefficiencies in the status quo, which requires three letters of recommendation for every student seeking admission to a graduate program. We suggest that instead students may submit a portfolio, highlighting 1–2 work products that can be directly assessed by admissions committees. These highlighted products would be accompanied by brief context statements submitted by a advisor, teacher, or mentor. The context statements will be qualitatively different from reference letters, as they will be very concise (we suggest limiting them to 1000 characters) and are focused on the student's work in creating the specific portfolio highlight. Our belief is that this new process will improve over the current situation, as it will be:

- Easier for faculty: the context statements will be submitted to just one platform, with a single simple form and a direct prompt requesting a brief response. Only 1–2 advisors will be asked for context statements, rather than 3 letter writers.
- More directly assessing the student's research: rather than trusting in and working to interpret the subjective opinions of letter writers, the admissions committee can directly assess the student's work. The portfolio also avoids the "all or nothing" situation currently in play, where only students with published papers have a product to show during graduate admissions; more diverse products and projects in different stages of readiness will be able to be shown in the portfolio.
- Less open to bias: the prompt for the context statement is specific to the portfolio products, keeping discussion focused on what the student has accomplished, and less on how the letter writer feels about the student.

## Acknowledgements

We are grateful for support from Cottrell Scholars Collaborative grant CS-CSC-2023-004, funded by the Research Corporation for Science Advancement.